\documentclass[pra,aps,twocolumn,floatfix]{revtex4}
\usepackage{epsfig} 

\begin{document} 
\title{Propagation of coupled dark-state polaritons and  storage of light in a tripod medium}

\author{Stefan Beck and Igor E. Mazets} 
\affiliation{
Vienna Center for Quantum Science and Technology, Atominstitut, TU~Wien,~Stadionallee~2,~1020~Vienna,~Austria; \\
Wolfgang Pauli Institute c/o Fakult\"{a}t f\"{u}r Mathematik,
Universit\"{a}t Wien, Oskar-Morgenstern-Platz 1, 1090 Vienna, Austria}

\begin{abstract} 
We consider the slow light propagation in an atomic medium with a tripod level scheme. We show that the coexistence 
of two types of dark-state polaritons leads to the propagation dynamics, which is qualitatively different from that 
in a $\Lambda $-medium, and allows therefore 
for very efficient conversion of signal photons into spin excitations. This 
efficiency is shown to be very close to 1 even for very long 
signal light pulses, which 
could not be entirely compressed into a $\Lambda $-medium at a comparable strength of the control field.  
\end{abstract}
\maketitle 

\section{Introduction} 
\label{s-i} 

The phenomenon of the electromagnetically induced transparency (EIT) based on the creation of a coherent superposition 
of long-living quantum states in  a medium irradiated by a two laser light fields has been known since a long ago, 
see, e.g., the review  
\cite{HarrisPT}. The EIT has become especially interesting and promising for the quantum memory applications 
\cite{qm} since 
the discovery of the method to ``stop the light" by conversion of photons of the weak (signal) 
into spin excitations of a medium by adiabatic turning off the second (control) field \cite{LF1,LF2}. 
Experimental realizations followed the publication of the idea \cite{LF1} 
immediately and employed as the EIT medium ultracold atoms \cite{Liu}, 
hot atomic vapor in a cell \cite{Phillips}, and doped crystal \cite{Turukhin}. 

\begin{figure}[b] 

\centerline{\epsfig{file=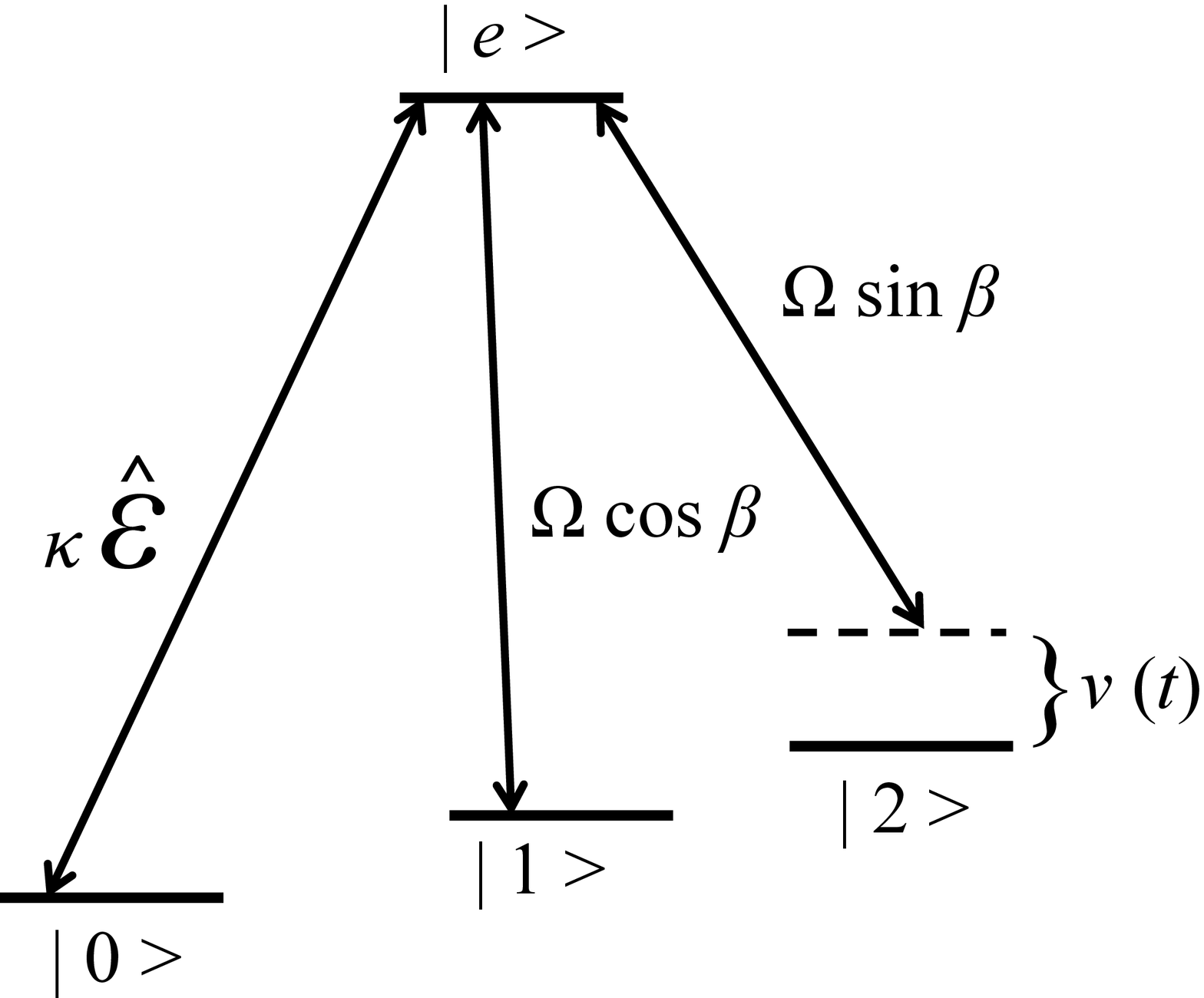,width=0.75\columnwidth }}

\begin{caption}
{Tripod scheme of atomic levels. 
\label{au-1} }
\end{caption}
\end{figure}

The $\Lambda $-scheme containing three quantum levels coupled to a laser radiation is a simplest one that admits 
coherent population trapping and the light propagation in a medium in the EIT regime. The tripod scheme that 
contains three low-energy, stable (or metastable) sublevels supports two different  quantum superpositions 
that are decoupled from coherent three-component laser radiation resonant to the optically excited state. 
Various aspects of the slow-light propagation and storage in a tripod medium have been 
theoretically studied \cite{Paspalakis,Petrosyan,Mazets2005,R1,gradient-sto,log-op,Ruseckas}. There are numerous 
experimental studies of the EIT in media with the tripod level scheme 
\cite{tripod2000e,Karpa,tripod2009e,tripod2011e,tripod2011ee,tripod2014e}, culminating in the demonstration 
of the storage and retrieval of light pulses at a single-photon level \cite{Pan}. 

The dynamical coupling between dark-states potations of the two types arising due to the time dependence 
of the control fields has been introduced in Ref. \cite{R1} but not fully investigated.
Indeed, the Hong-Ou-Mandel interferometer operation in a tripod medium \cite{R1} requires the change of 
the control field parameters during the time interval of no signal photon coming. In this paper we consider the 
situation of signal photons interacting with a tripod medium where the coupling between the two types 
of dark-state polaritons is present. 

The signal laser pulse can be fully converted into spin excitations in a conventional $\Lambda $-medium 
only if the medium is long enough to 
accommodate the whole slowed-down incoming pulse (spatially compressed in proportion to the ratio of its 
group velocity in the EIT regime and in the vacuum) \cite{Liu,Phillips,Turukhin}. 
In a tripod medium the existence of two coupled dark-state modes allows
for a conversion of almost the whole signal light field into spin
excitations under less restrictive conditions. 
Note that in a similar case a conventional $\Lambda $-type medium irradiated by
a control field of comparable strength and characterized by a 
comparable slow group velocity would accommodate only a part of
the signal pulse.

\section{Basic equations} 
\label{s-ii} 

We consider a medium consisting of atoms with the level scheme shown in Fig.~\ref{au-1}. The ground-state 
sublevels $|0\rangle $,  $|1\rangle $, and  $|2\rangle $
are coupled to an excited state  $|e\rangle $ by three coherent fields. The control fields driving the 
transitions  $|1\rangle  \leftrightarrow |e\rangle $ and $|2\rangle  \leftrightarrow |e\rangle $ are 
characterized by Rabi frequencies $\Omega _1 \equiv \Omega \cos \beta $ and $\Omega _2 \equiv \Omega \sin \beta $, 
respectively. These fields are phase-locked or obtained from a common source by acousto-optical modulator in 
order to provide perfect cross-correlation of their noise and to prevent thus a noise-induced decay of the 
quantum coherence between the states $|1\rangle $ and $ |2\rangle $ \cite{Dalton}.  
The transition $|0\rangle  \leftrightarrow |e\rangle $ is driven by a quantized signal field. 
We can consider a field propagating freely through the atomic sample \cite{Pan} or 
in a nanofiber \cite{Arno1,Kimble2012}; the propagation direction of the signal field defines the axis $z$. 
The quantum field for the signal photons can 
be expressed as $\hat{\cal E}(z,t)\exp [-i\omega _{e0}(t-z/c)]$, where 
$\hat{\cal E}$ is its slowly varying amplitude subjected to the standard bosonic commutation rules, 
$\omega _{e0}$ is the resonance frequency of the  $|0\rangle  \leftrightarrow |e\rangle $ atomic transition, 
$c$ is the speed of light (in vacuum or in the nanofiber, depending on the type of the set-up). 

In contrast to Ref. \cite{R1}, we assume the amplitudes of the two  control fields to be constant, but instead 
we introduce the detuning of the second field $\nu (t)$, which is time-dependent in a general case, see 
Fig.~\ref{au-1}. It is convenient to express the atomic collective spin variables through bosonic fields 
$\hat f_\alpha (z,t)$, where $\hat f_\alpha (z,t)$ annihilates an atom in the state 
$|\alpha \rangle $, $\alpha =0,1,2,e$, at the point $z$ at time $t$. The set of equations for these bosonic fields 
and the signal photons is 
\begin{eqnarray} 
\frac \partial {\partial t}\hat {\cal E }&=& -c\frac \partial {\partial z}\hat {\cal E }+i\kappa 
\hat f_0^\dag  \hat f_e, \label{e-E}\\
\frac \partial {\partial t}\hat {f}_0 &=& i\kappa \hat {\cal E}^\dag \hat f_e ,    \label{e-f0} \\
\frac \partial {\partial t}\hat {f}_e &=&i\kappa \hat{\cal E}\hat f_0+i\Omega (\cos \beta \hat f_1 +
\sin \beta \hat f_2),  \label{e-fe} \\
\frac \partial {\partial t}\hat {f}_1&=& i\Omega \cos \beta \hat f_e  ,   \label{e-f1} \\ 
\frac \partial {\partial t}\hat {f}_2&=&i\nu (t)\hat f_2+i\Omega \sin \beta \hat f_e  , \label{e-f2}
\end{eqnarray} 
which is an obvious generalization from the case of EIT in a $\Lambda $-medium \cite{Mazets2014} 
to the case of tripod medium. The atom-field coupling constant $\kappa = 
d_{e0} \sqrt{\omega _{e0}/(2\hbar \varepsilon _0A) } $, where $d_{e0}$ is the electric dipole moment of the 
transition $|0\rangle  \leftrightarrow |e\rangle $ and $A$ is the effective area of the signal beam, can be 
expressed through the optical density $s$ of the medium for the resonant signal light as 
$\kappa =\sqrt{ \gamma s c/(2N)}$, where $N$ is the number of atoms inside the interaction volume $AL$ and 
$L$ is the atomic sample length in the propagation direction. The radiative decay rate $\gamma $ arises 
due to the coupling of the $|0\rangle  \leftrightarrow |e\rangle $ transition to side modes of the 
electromagnetic field, which is not explicitly written, for the sake of brevity, in Eq. (\ref{e-fe}). 
Integrating out the vacuum modes of the electromagnetic field, we would get, instead of Eq. (\ref{e-fe}), 
\begin{equation} 
\frac \partial {\partial t}\hat {f}_e =i\kappa \hat{\cal E}\hat f_0+i\Omega (\cos \beta \hat f_1 +
\sin \beta \hat f_2)-\gamma \hat f_e +\hat \varsigma _e(z,t) ,
\label{e-fe-g}
\end{equation}
where $\hat \varsigma _e(z,t)$ is a delta-correlated Langevin-type operator \cite{LF1,LF2} that  
describes vacuum quantum noise and is needed to preserve bosonic commutation properties of $\hat f_e$ after 
introducing the decay term $-\gamma \hat f_e$. 

We work in the weak-pulse limit, i.e., assume that the linear density of dark-state polaritons is always 
much less than the linear density of atoms $n_\mathrm{1D}=N/L$, 
which are initially all in the state $|0\rangle $ \cite{Mazets2014,Kuang}. Then we linearize 
Eqs. (\ref{e-E}--\ref{e-f2}) by replacing $\hat f_0$ by a number $\sqrt{n_\mathrm{1D}}$ and find in a standard way 
\cite{LF1,LF2,R1}, i.e., by adiabatic elimination of excitation modes separated from the 
dark-state polaritons by large energy gaps, the equations of motion 
\begin{eqnarray} 
\left( \frac \partial {\partial t}+v_\mathrm{g} \frac \partial {\partial z}\right) \hat \Psi &=& 
i\tilde \nu (t) \sin \tilde \beta (\sin \tilde \beta \hat \Psi +\cos \tilde \beta \hat \Upsilon ) , 
\label{e-X} \\
\frac \partial {\partial t}\hat \Upsilon &=& 
i\tilde \nu (t) \cos \tilde \beta (\sin \tilde \beta \hat \Psi +\cos \tilde \beta \hat \Upsilon ) 
\label{e-Y} 
\end{eqnarray}  
for two dark-state polariton fields 
\begin{eqnarray} 
\hat \Psi &=& \cos \theta \hat{\cal E}-\sin \theta (\cos \beta \hat f_1+\sin \beta \hat f_2) , 
\label{def-X}  \\
\hat \Upsilon &=& \sin \beta \hat f_1-\cos \beta \hat f_2. 
\end{eqnarray}  
The mixing angle in Eq. (\ref{def-X}) is defined by the usual expression 
$\tan \theta =\kappa \sqrt{n_\mathrm{1D}}/\Omega $, and $v_\mathrm{g}=c \cos ^2 \theta $ \cite{LF1,LF2}. Also we get 
$\tilde \nu (t) =(\sin ^2\theta \sin ^2\beta +\cos ^2 \beta )\nu (t)$ and $\tan \tilde \beta =\sin \theta 
\tan \beta $. 

We assume the slow-light regime, $\Omega \ll \kappa \sqrt{n_\mathrm{1D}}$ and, hence, $\sin \theta \approx 1$, 
$v_\mathrm{g}\ll c$. In this limit, $\tilde \nu (t)\approx \nu (t)$ and $\tilde \beta \approx \beta $. 
In what follows, we do not distinguish therefore between the values with and without tilde 
and omit this  symbol over $\nu $ and $\beta $. 

Before specifying the initial and boundary conditions to Eqs. (\ref{e-X}, \ref{e-Y}), we reformulate them 
for classical complex variables $\Psi $ and $\Upsilon $. This may be done for coherent states of the dark-state 
polariton fields as well as for a single-quantum states. In the latter case, we use the Schr\"{o}dinger representation 
and write the wave function of the system as $|\Xi (t)\rangle =\int _0^L dz\, [\Psi (z,t)\hat \Psi ^\dag (z,t)+
\Upsilon (z,t)\hat \Upsilon ^\dag (z,t)]|\mathrm{vac}\rangle +|\Xi _\mathrm{ph}(t)\rangle $, where 
$|\mathrm{vac}\rangle $ is the vacuum state of excitations (all atoms being in their internal state $|0\rangle $), 
and $|\Xi _\mathrm{ph}(t)\rangle $ describes a single photon either before entering the medium (at $z<0$) or 
after leaving it (at $z>L$). The evolution of $|\Xi _\mathrm{ph}(t)\rangle $ is not interesting for us, and the 
evolution of the remaining component of $|\Xi (t)\rangle $ is given by Eqs. (\ref{e-X}, \ref{e-Y}) with 
the operators $\hat \Psi $ and $\hat \Upsilon $ replaced by the complex fields $\Psi $ and $ \Upsilon $, 
respectively. 

It is convenient to introduce new variables, $\tau =t-z/v_\mathrm{g}$ and $\zeta =z/v_\mathrm{g}$. 
Then $\nu (t) =\nu (\tau +\zeta )$ and the 
equations of motion for dark-state polaritons become 
\begin{eqnarray} 
\frac \partial {\partial \zeta } \Psi &=& 
i\nu   \sin  \beta (\sin  \beta \Psi +\cos   \beta \Upsilon ) , 
\label{e-Xzeta} \\
\frac \partial {\partial \tau } \Upsilon &=& 
i \nu   \cos \beta (\sin \beta  \Psi +\cos  \beta  \Upsilon ) . 
\label{e-Ytau} 
\end{eqnarray}  
The boundary and initial conditions are 
\begin{equation} 
\Psi (0,\tau ) =\Psi _0(\tau ), \qquad \Upsilon (\zeta ,0)=0, 
\label{bic} 
\end{equation} 
where the function $\Psi _0(\tau )$ is determined by the shape of the incoming signal 
light pulse. We assume 
that $\Psi _0(\tau )=0$ for $\tau \leq 0$.  

\section{Propagation dynamics} 
\label{sec-iii} 
The main features of the dark-polariton dynamics can be determined from the solution of 
Eqs. (\ref{e-Xzeta}--\ref{bic}) in the case of constant two-photon detuning, $\nu \equiv \nu _0 =\, \mathrm{const}$. 
A simple phase transformation 
\begin{eqnarray} 
\Psi (\zeta ,\tau )&=&e^{i\chi (\zeta ,\tau )}\Psi ^\prime (\zeta ,\tau ), \quad 
\Upsilon (\zeta ,\tau )=e^{i\chi (\zeta ,\tau )}\Upsilon ^\prime (\zeta ,\tau ), \nonumber \\
{\chi (\zeta ,\tau )}&=& \nu _0 (\sin ^2 \beta \, \zeta +\cos ^2 \beta \, \tau  ) 
\label{phase-tr} 
\end{eqnarray} 
casts Eqs. (\ref{e-Xzeta}--\ref{e-Ytau}) into the form 
\begin{eqnarray} 
\frac \partial {\partial \zeta } \Psi ^\prime &=& 
i\nu _0 \sin  \beta \, \cos   \beta \Upsilon ^\prime  , 
\label{e-Xprime} \\
\frac \partial {\partial \tau } \Upsilon ^\prime &=& 
i \nu _0  \sin \beta  \, \cos \beta \Psi ^\prime  . 
\label{e-Yprime} 
\end{eqnarray}
A similar set of equations has been derived in Ref. \cite{R1} for a different
driving protocol of the tripod medium where the coupling between the
the two dark-state polaritons was induced by changing the angle $\beta $ 
in time. We note that the influence of this coupling on the pulse
propagation was not studied there.

Eqs. (\ref{e-Xprime}, \ref{e-Yprime})  
can be easily solved by means of Laplace's transform. Also we can note that after elimination  
of one of the fields the equation for the remaining one is reduced to the relativistic Klein-Gordon equation 
\begin{eqnarray} 
\frac \partial {\partial \zeta }\frac \partial {\partial \tau } \Psi ^\prime &=& 
\frac 14  \left( \frac {\partial ^2}{\partial T^2 } -
\frac {\partial ^2}{\partial X^2 }\right) \Psi ^\prime \nonumber \\ 
&=&-(\nu _0  \sin \beta  \, \cos \beta )^2\Psi ^\prime  , 
\label{KGe} 
\end{eqnarray} 
where 
$
\frac {\partial }{\partial T}\equiv \frac {\partial }{\partial \tau }+\frac {\partial }{\partial \zeta }
$  and $ 
\frac {\partial }{\partial X}\equiv \frac {\partial }{\partial \tau }-\frac {\partial }{\partial \zeta }
$.
Green's function for the Klein-Gordon equation are well known \cite{Greiner} 
and can be used to solve Eqs. 
(\ref{e-Xprime},~\ref{e-Yprime}). 

Finally, we obtain the solutions, 
\begin{widetext} 
\begin{eqnarray} 
\Psi (\zeta ,\tau )&=& e^{i\nu_0\sin ^2\beta \, \zeta } \left[ \Psi _0(\tau ) - 
\nu _0 \sin \beta \cos \beta \int _0^\tau d\tau ^\prime \,   \Psi _0(\tau -\tau ^\prime ) 
e^{i\nu_0\cos ^2\beta \, \tau ^\prime } \sqrt{\frac \zeta {\tau ^\prime }}
J_1 (2\nu _0 \sin \beta \cos \beta \, \sqrt{\zeta {\tau ^\prime }})\right]  , 
\label{sol-Psi}       \\ 
\Upsilon  (\zeta ,\tau )&=&i\nu _0 \sin \beta \cos \beta \, e^{i\nu_0\sin ^2\beta \, \zeta }
\int _0^\tau d\tau ^\prime \,   \Psi _0(\tau -\tau ^\prime ) 
e^{i\nu_0\cos ^2\beta \, \tau ^\prime }J_0 (2\nu _0 \sin \beta \cos \beta \, \sqrt{\zeta {\tau ^\prime }}),  
\label{sol-Y}
\end{eqnarray} 
\end{widetext} 
where $J_0$ and $J_1$ are the Bessel functions of the zeroth and first order, respectively. 

First we analyze Eq. (\ref{sol-Psi}). The first term in its r.h.s. describes, apart from gaining a 
$\zeta $-dependent phase shift, pulse propagation at the group velocity $v_\mathrm{g}$. However, this regime 
holds only for small values of $\zeta $, where the integrand in the second term is small because of the small 
values taken by the Bessel function  $J_1 (2\nu _0 \sin \beta \cos \beta \, \sqrt{\zeta {\tau ^\prime }})$. 
This propagation picture is typical for dark-state polaritons in a $\Lambda $-medium. However, it will be distorted 
at larger distances, when the second term becomes important. What happens then, one can see from the analysis 
of the dynamics of the $\Upsilon $ field. 

This field corresponds to spin excitations, which possess zero group velocity and are induced by coupling to 
the $\Psi $-type polaritons through the finite two-photon detuning $\nu _0$. 
Eqs. (\ref{e-Xprime}, \ref{e-Yprime}) together with the initial condition $\Upsilon (\zeta ,0)=0$ yields a simple 
conservation law     
\begin{equation} 
\int _0^{\zeta _L}\! d\zeta \,  |\Upsilon (\zeta  ,\tau )|^2  =  
\int _0^\tau \! d\tau ^\prime  |\Psi (0,\tau ^\prime )|^2    
-\int _0^\tau \! d\tau ^\prime  |\Psi (\zeta _L ,\tau ^\prime )|^2, 
\label{c-law}
\end{equation} 
where $\zeta _L=L/v_\mathrm{g}$. It relates the total population of the $\Upsilon $-type mode inside the 
medium of the length $L$ to the loss of the output pulse energy at the exit from the medium compared to 
the case of propagation in a $\Lambda $-medium at the group velocity $v_\mathrm{g}$. We define the 
efficiency of the conversion of signal photons to $\Upsilon $-type spin excitations as 
\begin{equation}  
\eta (\tau  ) = \frac{ \int _0^{\zeta _L}\! d\zeta \,  |\Upsilon (\zeta  ,\tau )|^2 }{\int _0^\infty 
\! d\tau ^\prime  |\Psi (0,\tau ^\prime )|^2}.
\label{def-eta}
\end{equation}  

From Eq. (\ref{sol-Y}) we can evaluate the enumerator of Eq. (\ref{def-eta}) if we recall the formula for 
integral of a product of two Bessel functions \cite{AbrSteg1}:
\\
\begin{widetext}
\begin{eqnarray} 
\int _0^{\zeta _L} d\zeta \,  |\Upsilon (\zeta  ,\tau )|^2&=&
\nu _0\sin \beta \cos \beta \int _0^\tau d\tau _1 \int _0^\tau d\tau _2 \, 
\Psi _0(\tau _1)\Psi _0^*(\tau _2) e^{-i\nu _0\cos^2\beta (\tau _1 -\tau _2)} \frac 1{\tau _1 -\tau _2}
\nonumber \\*
&& \times \Big{ \{ }  \sqrt{\tau -\tau _2} J_0[2\nu _0\sin \beta \cos \beta \sqrt{ \zeta _L(\tau-\tau _1)}]
J_1[2\nu _0\sin \beta \cos \beta \sqrt{ \zeta _L(\tau-\tau _2) }]    \nonumber \\*  
&& -  \sqrt{\tau -\tau _1} J_0[2\nu _0\sin \beta \cos \beta \sqrt{ \zeta _L(\tau-\tau _2)}]
J_1[2\nu _0\sin \beta \cos \beta \sqrt{ \zeta _L(\tau-\tau _1) }]    \Big{ \} }  . 
\label{int-Y1}
\end{eqnarray} 
\end{widetext}

\begin{figure}[!b] 

\centerline{\epsfig{file=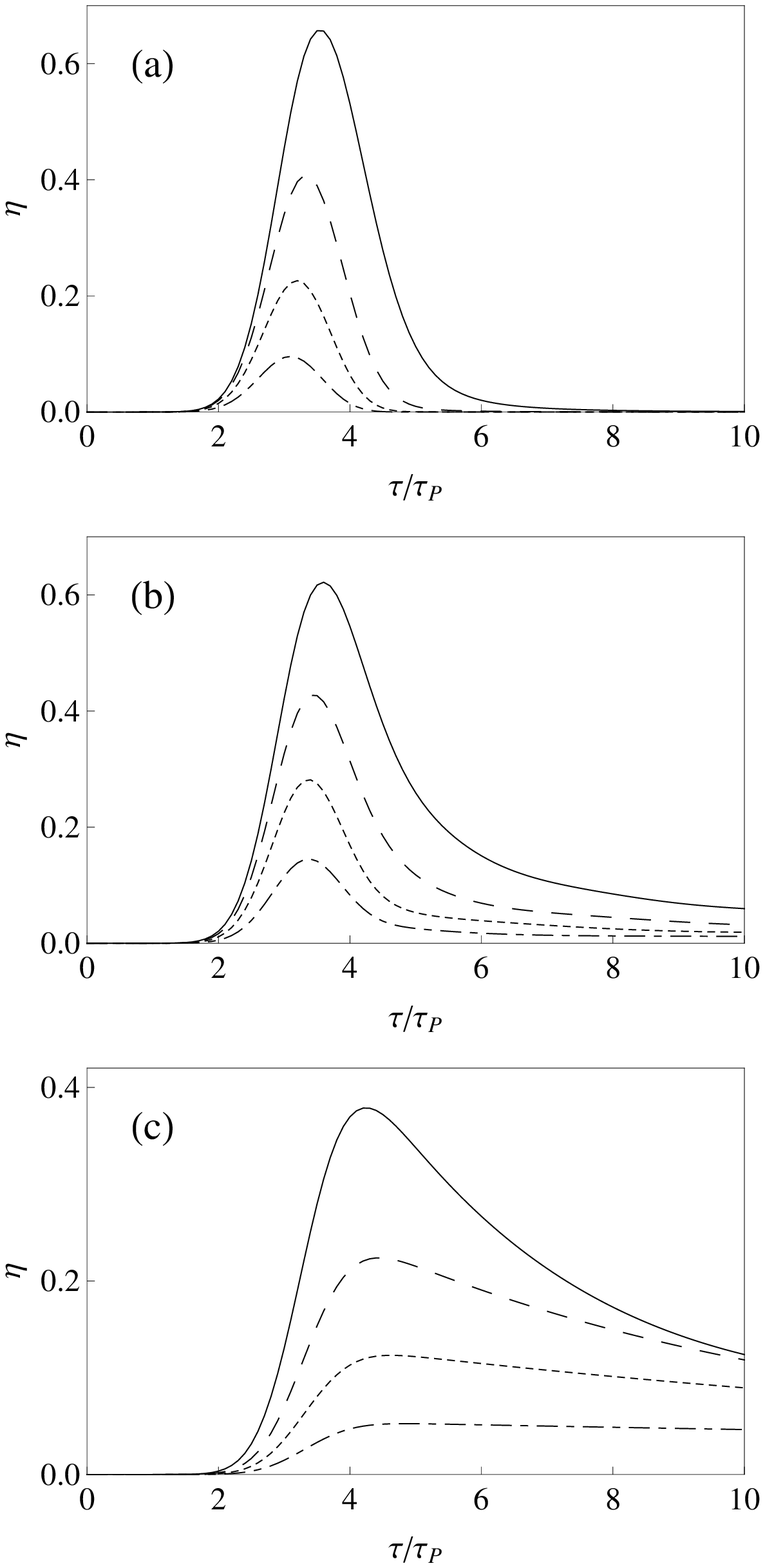,width=0.99\columnwidth }}

\begin{caption}
{Efficiency $\eta $ of the conversion of the signal photons into spin excitations of the $\Upsilon $-type as a 
function of the retarded time $\tau $ for the boundary 
condition $\Psi _0(\tau )=\Psi _{0\, \mathrm{A}} 
\{ \exp [ -(\tau -3\tau _\mathrm{p})^2/\tau _\mathrm{p} ^2]-
\exp [ -(\tau +3\tau _\mathrm{p})^2/\tau _\mathrm{p} ^2]\} $ for $\tau \geq 0$ ($|\Psi _{0\, 
\mathrm{A}}|^2$  determines the pulse energy). 
Units on the axes are dimensionless, the time 
$\tau $ is scaled to the characteristic time $\tau _\mathrm{p}$ of the pulse duration. $\beta =\pi /4$ 
for all plots. The detuning $\nu _0 $ is (a) $\nu_{0}=1/\tau_{p}$, (b) $5/\tau_{p}$, and (c) $10/\tau_{p}$.
On each panel the length $L$ of the medium is $1.0\, v_\mathrm{g}\tau _\mathrm{p}$ (solid line), 
$0.5\,  v_\mathrm{g}\tau _\mathrm{p}$ (long-dashed line), 
$0.25\, v_\mathrm{g}\tau _\mathrm{p}$ (short-dashed line), and  
$0.1\,  v_\mathrm{g}\tau _\mathrm{p}$ (dot-dashed line). The decrease of the dot-dashed line in panels (b) and (c) 
is very slow and can bee seen on a time scale $\tau /\tau _\mathrm{p} \sim 10^2$.  
\label{au-2} }
\end{caption}
\end{figure}

\begin{figure}[!b] 

\centerline{\epsfig{file=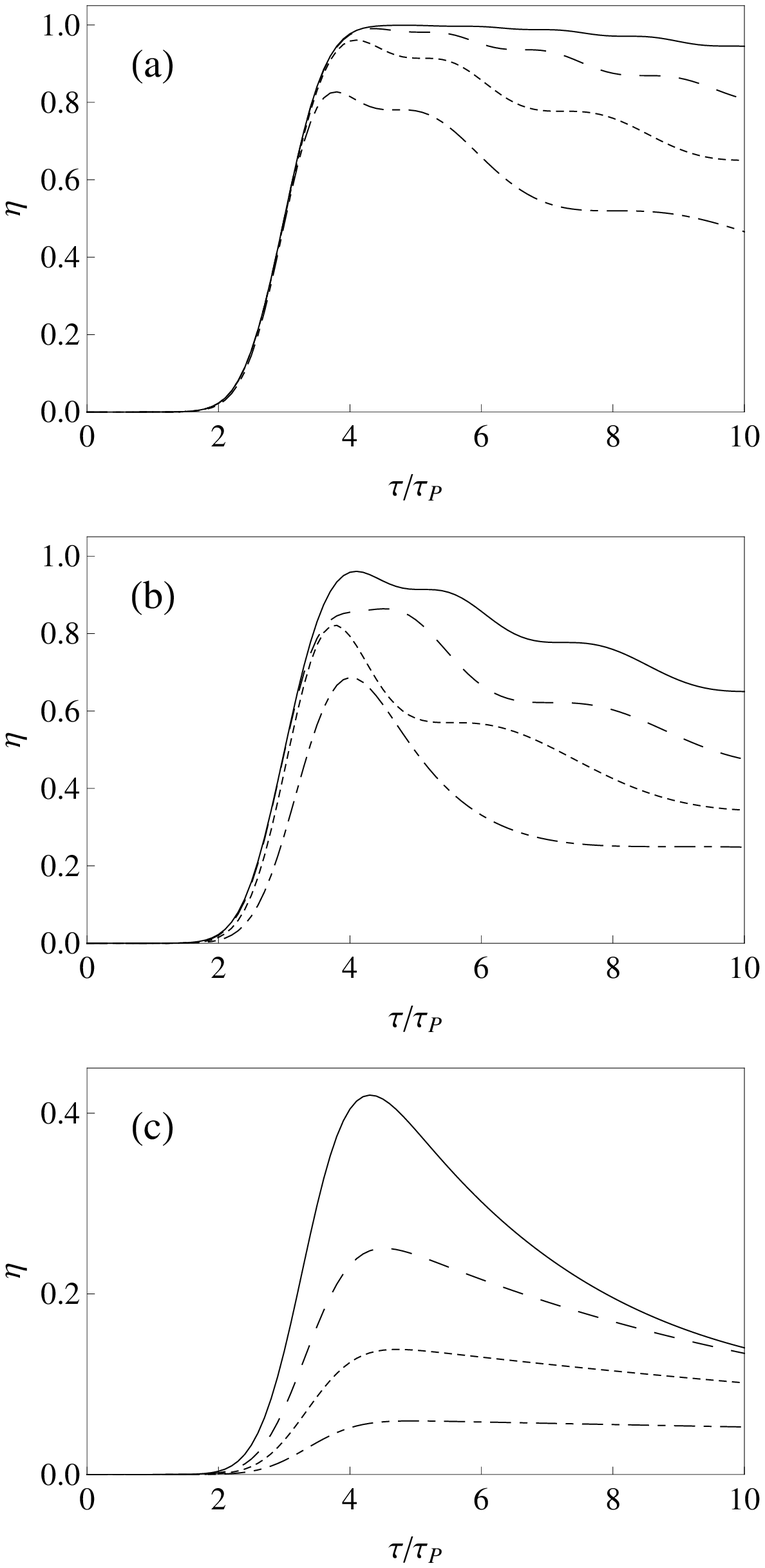,width=0.99\columnwidth }}

\begin{caption}
{The same as in Fig. \ref{au-2}, but for $\Psi _0(\tau )=\Psi _{0\, \mathrm{A}} 
\{ \exp [ -(\tau -3\tau _\mathrm{p})^2/\tau _\mathrm{p} ^2]-
\exp [ -(\tau +3\tau _\mathrm{p})^2/\tau _\mathrm{p} ^2]\}  \exp (i\nu _0\cos ^2\! \beta \, \tau )$ 
for $\tau \geq 0$.  
The values of $\nu _0$ assigned to the panels (a~--~c) and of $L$ assigned to the lines of different type as 
well as $\beta $ are the same as in the previous plot. Note that for a properly detuned signal pulse 
$\eta \approx 1$ can be attained. 
\label{au-3} }
\end{caption}
\end{figure}

Let $\tau _\mathrm{p}$ be the typical time scale of the incoming signal light pulse and consider Eq. (\ref{int-Y1}) 
for $\tau \gtrsim \tau _\mathrm{p}$. In this case the time integrals practically converge to their final values on 
the scale $\tau _{1,2} \lesssim \tau $. Assume that 
\begin{equation} 
|\nu _0\sin \beta \cos \beta |\sqrt{ \zeta _L\tau }\gg 1. 
\label{cond-1} 
\end{equation}
In this case we can recall the asymptotic expression \cite{AbrSteg2} 
for the Bessel function of the order $n$ of a large argument $x\rightarrow \infty $, 
$J_n(x) =\sqrt{2/(\pi x)}\cos ( x -\frac {n\pi }2-\frac \pi 4)$, and find an approximation for Eq. (\ref{int-Y1}) 
\begin{eqnarray} 
\int _0^{\zeta _L} d\zeta \,  |\Upsilon (\zeta  ,\tau )|^2=
\nu _0\sin \beta \cos \beta \int _0^\tau d\tau _1 \int _0^\tau d\tau _2 \, 
\Psi _0(\tau _1) && \nonumber \\ 
\qquad \times \Psi _0^*(\tau _2) e^{-i\nu _0\cos^2\beta (\tau _1 -\tau _2)} ~\qquad ~ &&  \nonumber \\ 
\qquad \times 
\frac {\sin [|\nu _0\sin \beta \cos \beta |\sqrt{ \zeta _L/\tau }\, (\tau _1-\tau _2)]}{\pi (\tau _1 -\tau _2)}. 
~~~~~ &&
\label{int-Y2}
\end{eqnarray}

If we consider asymptotically long times, such that 
\begin{equation} 
|\nu _0\sin \beta \cos \beta |\sqrt{ \zeta _L/\tau } \ll \frac 1{\tau _\mathrm{p}}, 
\label{cond-2}
\end{equation} 
we immediately see that asymptotically 
\begin{eqnarray}
\int _0^{\zeta _L} d\zeta \,  |\Upsilon (\zeta  ,\tau )|^2\vert _{\tau \rightarrow +\infty }  \approx 
\frac {\nu _0 \sin \beta \cos \beta }\pi \sqrt{ \frac {\zeta _L}\tau }  &&
\nonumber \\ 
~\qquad \times \left| \int _0^\infty d\tau ^\prime \, \Psi _0 (\tau ^\prime ) 
e^{-i\nu_0 \cos ^2 \beta \, \tau ^\prime } \right| ^2 .  
&& 
\label{int-Y3} 
\end{eqnarray} 
This means that the spin excitations in the medium, being coupled to the $\Psi $-type dark-state polariton mode 
via the two-photon detuning $\nu _0$, decay in a very slow, non-exponential way, namely, proportionally 
to $1/\sqrt{\tau }$.  

What occurs for $\tau $ larger than but close to $\tau _\mathrm{p}$, requires further analysis. 
Recall that we are interested in long pulses, which cannot be entirely fit into the medium. Therefore we 
assume $\tau _\mathrm{p}\gtrsim \zeta _L=L/v_\mathrm{g}$. 
Also we need, in order to satisfy the condition (\ref{cond-1}), 
to have values of $\beta $ not too close to $0, ~\pm \frac \pi 2$, or $\pi $. In other words, we assume that 
$\cos \beta $ and $\sin \beta $ are of the same order.
If the incoming signal photons are tuned exactly in resonance with the $|0\rangle \leftrightarrow |e\rangle $ 
transition, then $|\nu _0\sin \beta \cos \beta |\sqrt{ \zeta _L/\tau } \lesssim |\nu _0|\cos ^2 \beta $,   
the convergence of the integrals in Eq. (\ref{int-Y2}) is achieved on the time scale of about 
$1/(|\nu _0|\cos ^2 \beta )$ and Eq. (\ref{int-Y3}) remains a satisfactory estimation. The conversion 
efficiency $\eta $ thus remains well below 1. However, if the detuning of the 
signal pulse from the resonance is chosen such that $\Psi _0(\tau )=|\Psi _0 (\tau )| \exp (i\nu _0 \cos ^2\beta \, 
\tau )$ then, in order to determine the time scale of convergence of the time integrals in Eq. (\ref{int-Y2}), we 
have to compare $|\nu _0\sin \beta \cos \beta |\sqrt{ \zeta _L/\tau } \sim 
|\nu _0\sin \beta \cos \beta |\sqrt{ \zeta _L/\tau _\mathrm{p} } $ with the spectral width of $|\Psi _0|$, i.e., 
with $1/\tau _\mathrm{p}$. By taking $|\nu _0|$ large enough, one may attain 
$|\nu _0\sin \beta \cos \beta |\sqrt{ \zeta _L \tau _\mathrm{p} } \gg 1$. In this case, 
the function $\sin [|\nu _0\sin \beta \cos \beta |\sqrt{\zeta _L/\tau }(\tau _1-\tau _2)]/[\pi (\tau _1-\tau _2)]$ 
can be approximated by a delta-function, $\delta (\tau _1-\tau _2)$. Therefore the efficiency of conversion 
of signal photons into spin excitations of the $\Upsilon $-type is 
\begin{equation}  
\eta (\tau  ) \approx    
\frac{ \int _0^\tau  \! d\tau ^\prime  |\Psi (0,\tau ^\prime )|^2 }{\int _0^\infty 
\! d\tau ^\prime  |\Psi (0,\tau ^\prime )|^2}.
\label{extrem-eta}
\end{equation} 
For $\tau \gtrsim \tau _\mathrm{p}$ (say, $\tau \approx 3 \tau _\mathrm{p}$) this efficiency may get very close to 1. 
Of course, at very large times $\eta (\tau )$ decays, as we have shown, in proportion to $1/\sqrt{\tau }$. But 
one can prevent such a long-time decay of the spin excitations 
by radiating photons out of the medium 
by sudden switching off the control fields (or by sudden changing $\nu $ from $\nu _0$ to 0 and 
thus decoupling the $\Psi $-type and $\Upsilon $-type polaritons). This means that the use of a tripod medium permits 
one to trap and convert into spin excitations very long signal light pulses, which would be only partially fit into a 
medium with a standard $\Lambda $-scheme of atomic levels. The retrieval of the stored quanta may be implemented 
using the standard protocol \cite{R1} (note also the observed modulation of the retrieved pulse shape \cite{Pan}). For example, one may retrieve stored signal photons by applying the two control 
fields with the new amplitudes $\Omega _1^\mathrm{new}= \sin \beta \,  \Omega $ and 
$\Omega _2^\mathrm{new}=-\cos \beta \, \Omega $ and 
zero detuning, $\nu ^\mathrm{new}(t)=0$. Then the spin excitation stored in the medium turns into 
a dark-state polariton that moves at the group velocity $v_\mathrm{g}$ and leaves the medium without 
coupling to the dark-state polariton of the other type. 

The numerical evaluation of the efficiency of conversion of signal photons into spin excitations based on 
Eq. (\ref{sol-Y}) is presented in Figs.~\ref{au-2} and \ref{au-3}. The incoming pulse used here is 
slightly (at the level of $10^{-4}$) 
modified in comparison to a Gaussian in order to formally 
provide its continuity at $\tau =0$, since  
$\Psi _0(0)=0$ by our assumption. 
We can see from Fig.~\ref{au-2} 
that for a perfectly resonant 
signal light the maximum efficiency is always appreciably below 1. On the contrary, if the probe light is detuned 
by $-\nu _0\cos ^2 \beta $ from the frequency $\omega _{e0}$, than the values of $\eta $ very close to 1 can be 
attained, see Fig.~\ref{au-3}(a). Note that for the $L/(v_\mathrm{g}\tau _\mathrm{p})$ equal to 1.0 and 0.5 the 
maximum fraction of the Gaussian incident pulse that can be simultaneously contained inside the medium is 
0.843 and 0.521, respectively, while the maximum values of $\eta $ on the  Fig.~\ref{au-3}(a) for the 
respective lengths are 0.999 (solid line) and 0.990 (long-dashed line). 

\section{Discussion}
\label{sec-iv}  

Now we examine the effects of the signal light absorption on the propagation regime considered in the previous Section. 
Eq. (\ref{cond-1}) ensures highly efficient conversion of signal photons into spin excitations on intermediate 
time scales of about few incoming pulse duration times $\tau _\mathrm{p}$. This high efficiency implies 
a large optical density of the medium, $s\gg 1$. Also we consider $\beta \approx \frac \pi 4$. 
Since $v_\mathrm{g}\approx c\Omega ^2/(\kappa ^2 n_\mathrm{1D})\ll c$ and $\kappa ^2n_\mathrm{1D}L =
\gamma c s/2$, we can rewrite Eq. (\ref{cond-1}) as 
\begin{equation} 
|\nu _0 |\sqrt{ \frac {\sqrt{s} \tau _\mathrm{p}}{\Delta \omega _\mathrm{EIT}} }\gg 1, 
\label{cond-3} 
\end{equation} 
where 
\begin{equation} 
\Delta \omega _\mathrm{EIT} = \frac {\Omega ^2}{\gamma \sqrt{s}} 
\label{w-EIT} 
\end{equation} 
is the width of the EIT transmission window in an optically dense medium \cite{LF2,Gornyi} (see also 
the review \cite{review1}). To minimize the signal pulse absorption, we have to assume its duration to be much longer 
than $1/\Delta \omega _\mathrm{EIT}$: 
\begin{equation} 
\tau _\mathrm{p} =\frac {K_\mathrm{p}}{\Delta \omega _\mathrm{EIT}}, \qquad K_\mathrm{p}\gg 1. 
\label{def-K} 
\end{equation}
Also the two-photon detuning must be small compared to the width of the EIT window, $|\nu _0|\ll 
\Delta \omega _\mathrm{EIT}$. This means that for a large optical depth and long enough pulses, 
\begin{equation} 
\sqrt{s}K_\mathrm{p} \gg \left| \frac {\Delta \omega _\mathrm{EIT}}{\nu _0}\right| ^2 , 
\label{cond-4}
\end{equation} 
the condition (\ref{cond-1}) is satisfied. 


To summarize, we investigated theoretically the propagation of weak signal pulses in a medium with a tripod 
scheme of atomic levels in the slow-light regime. 
Dark-state polaritons of two kinds exist in such a medium \cite{R1}. When they are mutually coupled 
via non-zero detuning $\nu (t)$ of one of the control fields, the propagation becomes non-trivial. In the case 
of constant detuning, $\nu (t)\equiv \nu _0$, the propagation bears analogy with the relativistic physics because 
its Green's function is formally identical to that of the Klein--Gordon equation \cite{Greiner}. The light pulse leaving  
the medium has a very long ``tail" decreasing as $1/\sqrt{\tau }$. Under certain conditions [Eq. (\ref{cond-4}) 
together with the detuning of the signal pulse by $-\nu _0 \cos ^2\beta $ from the single-photon resonance] it is 
possible to trap temporally almost the entire incoming pulse even if it is so long that it cannot be accommodated in 
a $\Lambda $-medium characterized by a comparable reduction of the group velocity, $v_\mathrm{g}/c = 
\Omega ^2/(\kappa ^2n_\mathrm{1D})\ll 1$. Fast switching off the control fields or setting $\nu (t)$ to zero 
prevents the spin excitations from decay through the radiation of photons in the forward direction and leads 
to their storage in the medium. 
Note that the proposed scheme may be termed ``passive", since it, unlike the conventional one 
\cite{LF1,LF2,Liu,Phillips,Turukhin}, does not require gradual tuning of the control  field 
during the signal pulse propagation in the medium, but 
implies instead rapid switching off of both the control fields or a fast change of $\nu (t)$ to zero, 
as soon as the maximum conversion efficiency is achieved. 

\begin{acknowledgments}
The authors thank N.J. Mauser, A. Rauschenbeutel,  and Ph. 
Schneewei{\ss } for helpful discussions.
We acknowledge financial support
by the Austrian Ministry of Science, Research and Economy  (BMWFW) 
via its grant for the Wolfgang Pauli Institute, 
by the EU via  the ERC advance grant QuantumRelax, 
and by the Austrian Science Foundation
(FWF) via the projects P~25329-N27, 
F41 (SFB ViCoM), and W1245 (DK ``Dissipation und Dispersion 
in nichtlinearen partiellen  Differentialgleichungen").
\end{acknowledgments}

\end{document}